\documentstyle[multicol,epsf,aps]{revtex}
\begin{document}
\title{Particle Systems with Stochastic Passing}
\author{I. Ispolatov$^1$ and P. L. Krapivsky$^2$} 
\address{$^1$Department of Chemistry , Baker Laboratory, Cornell University,
  Ithaca, NY 14853}
\address{$^2$Center for Polymer Studies and   
  Department of Physics, Boston University, Boston, MA 02215}
\maketitle 
 
\begin{abstract} 
  We study a system of particles moving on a line in the same direction.
  Passing is allowed and when a fast particle overtakes a slow particle,
  it acquires a new velocity drawn from a distribution $P_0(v)$, while
  the slow particle remains unaffected.  We show that the system reaches
  a steady state if $P_0(v)$ vanishes at its lower cutoff; otherwise, the
  system evolves indefinitely.

\end{abstract}

\begin{multicols}{2} 

\noindent    
We are interested in behavior of a system of interacting particles
moving on the real line in one direction, say to the right. The system
is endowed with the following simple dynamics: (i) Particles move freely
between ``collisions''; (ii) After a collision, or ``passing'' event,
the velocity of the slow particle remains the same, $v_{\rm
  slow}$=const, while the fast particle instantaneously acquires some
new velocity, $v_{\rm new}>v_{\rm slow}$, drawn from the intrinsic
velocity distribution $P_0(v)$.  We would like to answer the following
basic questions about the behavior of the system: Does the velocity
distribution $P(v,t)$ reach a steady state or the system continues to
evolve indefinitely? How does the average velocity depend on time? etc.
  
Our motivation is primarily conceptual, as we want to understand
non-equilibrium infinite particle systems with two-body interactions.
Thus we have chosen the simplest dynamics -- interactions occur only
upon colliding, and only one particle is affected.  The appealing
simplicity of the model suggests that it might show itself in different
natural phenomena, and indeed, we originally arrived to this model in an
attempt to mimic traffic on one-lane roads.  Somewhat related dynamics
were already used in modeling voting systems\cite{melzak,ff}, force
fluctuations in bead packs\cite{c}, asset exchange processes\cite{sps},
combinatorial processes\cite{d}, continuous asymmetric exclusion
processes\cite{kg}, granular gases\cite{in}, and
aggregation-fragmentation processes\cite{maj}.
  
Let us first consider discrete velocity distributions.  Specifically,
we assume that both initial velocities and new velocities are drawn from
the same intrinsic distribution $P_0(v)=\sum p_j\delta(v-v_j)$.  For the
binary distribution, the system does not evolve at all, so the first
non-trivial case is the ternary intrinsic distribution when the system
contains slow, moderate, and fast particles.  Initially,
\begin{equation}
\label{pv}
P_0(v)=p_1\delta(v-v_1)+p_2\delta(v-v_2)+p_3\delta(v-v_3).
\end{equation}
We set
\begin{equation}
\label{cond}
p_1+p_2+p_3=1, \quad
v_1<v_2<v_3,
\end{equation}
without a loss of generality. When the steady state is reached, the
velocity distribution remains ternary,
\begin{equation}
\label{qv}
P_{\rm eq}(v)=p_1\delta(v-v_1)+q_2\delta(v-v_2)+q_3\delta(v-v_3).
\end{equation}
The density of slow particles does not change, while the densities $q_2$
and $q_3$ of moderate and fast particles differ from the initial values.
The final densities are found from a simple probabilistic argument based
on the requirement of stationarity.  For moderate particles we get
\begin{equation}
\label{q2}
p_3(v_2-v_1)q_2=p_2(v_3-v_1)q_3.
\end{equation}
The left-hand side of Eq.~(\ref{q2}) gives the loss in $q_2$ which
happens when a moderate particle overtakes a slow particle and becomes a
fast particle.  The right-hand side gives the gain in $q_2$ which takes
place when a fast particle overtakes a slow particle and converts into a
moderate particle.  Solving Eq.~(\ref{q2}) together with the
normalization condition, $p_1+q_2+q_3=1$, we find
\begin{equation}
\label{q23}
q_2=p_2\,{p_2+p_3\over p_2+\nu p_3}, \quad
q_3=\nu p_3\,{p_2+p_3\over p_2+\nu p_3}.
\end{equation}
where $\nu=(v_2-v_1)/(v_3-v_1)$. Since $\nu<1$, we have $q_2>p_2$ and
$q_3<p_3$.  Thus the density of moderate particles increases while the
density of fast particles decreases.  Similarly, one can analyze
discrete velocity distributions with more than three particle species.
In all cases (i) the system reaches a steady state; (ii) the average
velocity decreases and eventually reaches some finite value; (iii) the
density of the most slow particle species remains unchanged.

Now we consider a continuous velocity distributions.  Let $[v_{\rm
  min},v_{\rm max}]$ be a support of the intrinsic velocity
distribution $P_0(v)$.  By Galilean transform, we can 
set $v_{\rm min}=0$ without loss of
generality.  We consider unbounded
distributions, $v_{\rm max}=\infty$, although main results equally apply
to the cases with finite $v_{\rm max}$.

The passing rule asserts that when a fast particle overtakes a slow
particle moving with a velocity $v_{\rm slow}$, the assignment of the new
velocity $v$ occurs with probability
\begin{equation}
\label{rule}
P_0\left(v|v_{\rm slow}\right)=P_0(v)\,
{\theta(v-v_{\rm slow})\over \int_{v_{\rm slow}}^\infty dv'\,P_0(v')}. 
\end{equation}
Eq.~(\ref{rule}) guarantees that $v>v_{\rm slow}$ and that the normalization
requirement, $\int dv\,P_0\left(v|v_{\rm slow}\right)=1$, is obeyed. 

Now we can write a Boltzmann equation for the velocity distribution 
$P(v,t)$:
\begin{eqnarray}
\label{main}
&&{\partial P(v,t)\over \partial t}=-P(v,t) \int_0^v dv' (v-v')P(v',t)\\
&&+\int_0^v dv_2\,P_0(v|v_2)
\int_{v_2}^\infty dv_1\,(v_1-v_2)P(v_1,t)P(v_2,t).\nonumber
\end{eqnarray}
The first term on the right-hand side of Eq.~(\ref{main})
describes loss in  $P(v,t)$ due to collisions with slow particles: 
Collisions occur with
rate proportional to velocity difference, and the integration limits
ensure that only collisions with slower particles are taken into
account. The second, gain, term accounts for the increase of $P(v,t)$
due to a random assignment of velocity $v$ after collision.

We could not solve Eq.~(\ref{main}) in the general case of an arbitrary
intrinsic velocity distribution $P_0(v)$.  Attempts to find a solution
even for some particularly simple $P_0(v)$, e.g., linear, exponential, or
uniform, turned out to be fruitless as well. Thus we proceed by
employing asymptotic, approximate, and numerical techniques.

We start by looking at the asymptotic behavior of $P(v)$ in the small
velocity limit. Let $v\ll u(t) $, where $u(t)$ is the average velocity,
\begin{eqnarray}
\label{avv}
u(t)\equiv \langle v\rangle=\int_0^\infty dv\,vP(v,t).
\end{eqnarray}
Then Eq.~(\ref{main}) simplifies to
\begin{eqnarray}
\label{asymp}
{\partial P(v,t)\over \partial t}&=& 
P_0(v)\,u(t)\int_0^v dv_2\,P(v_2,t)\nonumber\\
&-&P(v,t) \int_0^v dv' (v-v')P(v',t).
\end{eqnarray}
To probe the small $v$ behavior, we need to know 
$P_0(v)$ at $v\to 0$.  Let us consider a family of intrinsic velocity
distributions that behave algebraically:
\begin{equation}
\label{alg0}
P_0(v)\simeq Av^\mu \quad {\rm when}\quad 
v\to 0.
\end{equation}

Now {\em assume} that the system reaches the steady state: $P(v,t)\to
P_{\rm eq}(v)$ and $u(t)\to u_{\rm eq}$. Plugging these and
Eq.~(\ref{alg0}) into Eq.~(\ref{asymp}), we find that the velocity
distribution also behaves algebraically in the small velocity limit,
\begin{equation}
\label{alg}
P_{\rm eq}(v)\simeq (\mu+1)A\,u_{\rm eq}\, v^{\mu-1} \quad {\rm when}\quad 
v\to 0.
\end{equation}
In other words, the steady state velocity distribution scales as
$v^{-1}P_0(v)$.  Recalling the normalization requirement,
$\int_0^{\infty} dv\,P(v)=1$, we see that it is
possible only when $\mu>0$.  Thus our {\em assumption} that the system
reaches a steady state is certainly wrong when $\mu\leq 0$. In this
region, we anticipate that the system will evolve indefinitely. Note
that both the exponential and uniform intrinsic distributions belong to
the borderline case of $\mu=0$ that separates stationary and
evolutionary regimes; for them an anomalously slow kinetics is
anticipated.

To probe the behavior of evolving systems, we {\em assume} that in the
long-time limit there is a very small fraction of ``active'' particles
that move with velocities $v\sim 1$ and the vast majority of ``creeping''
particles that hardly move at all.  We ignore collisions between active
particles since their density is very low. We also ignore collisions
between creeping particles since their relative velocity is very small.
This picture suggests that only collisions between active and creeping
particles matter. Thence, the velocity distribution of active particles
obeys
\begin{equation}
\label{linear}
{\partial P(v,t)\over\partial t}=P_0(v)u(t)-vP(v,t).
\end{equation}
Eq.~(\ref{linear}) may at best describe the evolution process in the
long-time limit. However, for the sake of tractability, we apply it to
the whole time range and use the natural initial condition
$P(v,0)=P_0(v)$.  Eq.~(\ref{linear}) is an inhomogeneous linear
differential equation which is easily solved to give
\begin{equation}
\label{sol}
P(v,t)=P_0(v)\,e^{-vt}\left[1+\int_0^t dt'\, u(t')\,e^{vt'}\right].
\end{equation}
This solution implies
\begin{equation}
P(v,t)\sim  u(t)v^{-1} P_0(v) \qquad {\rm for}\qquad v\gg {1\over t},
\end{equation}
which resembles Eq.~(\ref{alg}).  

To close the solution of Eq.~(\ref{sol}), we must determine $u(t)$. It
is possible to plug (\ref{sol}) into the definition of the average
velocity, Eq.~(\ref{avv}), and get an integral equation for $u(t)$. In
the following we use another approach, which is technically simpler.
Note that the density of active particles, $\int dv\,P(v,t)$, is
manifestly conserved by Eq.~(\ref{linear}). After integration over
velocity, Eq.~(\ref{sol}) becomes
\begin{equation}
\label{cond1}
1=\hat P_0(t)+\int_0^t dt'\, u(t')\,\hat P_0(t-t'),
\end{equation}
where $\hat P_0$ is the Laplace transform of the intrinsic velocity
distribution,
\begin{equation}
\label{lap}
\hat P_0(t)=\int_0^\infty dv\, P_0(v)\,e^{-vt}.
\end{equation}

One can guess the long time behavior of $P(v)$ without actually solving
Eq.~(\ref{cond1}). Let us assume that the average velocity varies slowly
with $t$.  Then the integral on the right-hand side of Eq.~(\ref{cond1})
can be estimated as $u(t)\int_0^t dt'\,\hat P_0(t')$, which implies
\begin{equation}
\label{est}
u(t)\sim \left[\int_0^t dt'\,\hat P_0(t')\right]^{-1}.
\end{equation}
For an intrinsic velocity distribution with an algebraic behavior
(\ref{alg0}) in the small-$v$ limit, we have $\hat P_0(t)\sim
t^{-1-\mu}$ for large $t$.  Hence $\int_0^t dt'\,\hat P_0(t')\sim
t^{-\mu}$ for $\mu<0$, and it follows from Eq.~(\ref{est}) that
$u(t)\sim t^\mu$.  The above derivation is quite careless, though the
final result is correct.  Now we derive this result in a more rigorous
way.

The convolution form of the integral in Eq.~(\ref{cond1}) suggests to
apply the Laplace transform once more. It yields
\begin{equation}
\label{simple}
{1\over s}=\int_0^\infty dv\, {P_0(v)\over s+v}
+\hat u(s)\int_0^\infty dv\, {P_0(v)\over s+v}.
\end{equation}
Here
\begin{equation}   
\label{lapu}
\hat u(s)=\int_0^\infty dt\, u(t)\,e^{-st}
\end{equation}
is the Laplace transform of the average velocity.  The double Laplace
transform of the intrinsic velocity distribution has been simplified:
\begin{equation}
\label{lap2}
\int_0^\infty dt\,e^{-st}\int_0^\infty dv\, P_0(v)\,e^{-vt}=
\int_0^\infty dv\, {P_0(v)\over s+v}.
\end{equation}
Thus the Laplace transform of the average velocity is
\begin{equation}
\label{exact}
\hat u(s)=-1+\left[s\int_0^\infty dv\, {P_0(v)\over s+v}\right]^{-1}.
\end{equation}
Generally, one cannot obtain more explicit results.  Given that the
above approach describes only the long-time asymptotics, let us focus on
this regime. To probe the long-time behavior, one should determine the
small $s$ asymptotics of $\hat u(s)$.  For algebraic intrinsic velocity
distributions (\ref{alg0}), the asymptotics of (\ref{lap2}) reads
\begin{eqnarray}
\label{lap5}
\int_0^\infty dv\, {P_0(v)\over s+v}\to 
As^\mu \int_0^\infty dw\, {w^\mu\over w+1}={A\pi\over \sin(-\pi\mu)}\,s^\mu.
\end{eqnarray}
This applies for $-1<\mu<0$ (the lower bound comes from the
normalization requirement, $\int dv\,P_0(v)=1$). Plugging (\ref{lap5})
into (\ref{exact}) yields
\begin{equation}
\label{us}
\hat u(s)\to {\sin(-\pi\mu)\over A\pi}\,s^{-1-\mu}
\quad {\rm for} \quad s\to 0,
\end{equation}
and by making the inverse Laplace transform, we finally arrive at
\begin{equation}
\label{ut}
u(t)\to {\sin(-\pi\mu)\over A\pi\Gamma(1+\mu)}\,t^\mu
\quad {\rm for} \quad t\to \infty.
\end{equation}
This result agrees with the asymptotics we naively derived earlier.

A special consideration is required for the borderline case of $\mu=0$.
For concreteness, consider the exponential intrinsic distribution,
$P_0(v)=\exp(-v)$.  Its double Laplace transform reads
\begin{eqnarray*}
\int_0^\infty dv\, {e^{-v}\over s+v}=e^s E_1(s),
\end{eqnarray*}
where
\begin{eqnarray*}
E_1(s)=\int_1^\infty dx\,{e^{-xs}\over x}
\end{eqnarray*}
is the exponential integral. As a result, Eq.~(\ref{exact}) becomes
\begin{equation}
\label{hus}
\hat u(s)=-1+{1\over se^s E_1(s)}.
\end{equation}
Using the well-known asymptotics of the the exponential
integral\cite{bo}, $E_1(s)=-\ln s -\gamma+{\cal O}(s)$,
(where $\gamma\cong 0.5772$ is Euler's constant), we transform
Eq.~(\ref{hus}) into
\begin{equation}
\label{hus2}
\hat u(s)=-{1\over s(\ln s+\gamma)}+{\cal O}\left({1\over \ln s}\right).
\end{equation}
Performing the inverse Laplace transform gives
\begin{equation}
\label{ut2}
u(t)\to {1\over \ln t}\quad {\rm for} \quad t\to \infty.
\end{equation}
To summarize, for the family of intrinsic velocity distribution with
algebraic behavior in the small $v$ limit (\ref{alg0}), our predictions
for the long-time asymptotics of the average velocity $u(t)$ are:
\begin{equation}
\label{av}
u(t)\sim\cases{{\rm const},  & for $\mu>0$;\cr
               (\ln t)^{-1}, & for $\mu=0$;\cr
               t^\mu,        & for $-1<\mu<0$.\cr}
\end{equation}

To check the validity of asymptotic predictions and, more generally, to
see if the mean-field theory is applicable at all, we perform molecular
dynamics simulations and solve the Boltzmann equation (\ref{main})
numerically. To sample distinct regimes predicted in (\ref{av}) we
consider the intrinsic velocity distribution
\begin{eqnarray}
\label{p_0}
P_0(v)={v^\mu e^{-v}\over \Gamma(\mu+1)}
\end{eqnarray}
with $\mu=1, 0, -1/2$.  

In molecular dynamics simulations, we place $N$ particles onto the ring
of length $L=N$ so that the average density is equal to one.  Most of our
simulations are performed for $N=5\cdot 10^4$ particles, but we also
simulated twice larger system and found no appreciable difference.
Initially, particle velocities are randomly drawn from the distribution
$P_0(v)$. The model is updated according to the collision-time-list 
algorithm suggested in Ref.\cite{sid}.

To solve the Eq.~(\ref{main}) numerically, we use Euler's time update
with both uniform and non-uniform grid; in the latter case, we take
$v_N=(N/ N_{max})^4 v_{max}$ velocity grid with $v_{max}=15$ and
$N_{max}=300 - 500$.  Integrals on the right-hand side of
Eq.~(\ref{main}) are calculated using the trapezoid rule; time increment
$\delta t=0.1$ was found to be suitable for all three $P_0(v)$.

The main conclusion is that results of molecular dynamics simulations
and numerical solutions of the mean-field equation are virtually
identical (see, e.g., Fig.~1). It confirms our assumption that the
system remains well-stirred and no appreciable spatial correlations
develop.  Additionally, Fig.~1 shows that for $P_0(v)=v\,e^{-v}$, the
approach of $P(v,t)$ to the steady state is non-uniform in velocity.
This is caused by the obvious fact that for any finite time, the
velocity distribution $P(v,t)$ must still vanish at the lower cutoff as
$P_0(v)$ does. In other words, the steady state (\ref{alg}) is reached
outside the ``boundary layer'', $v\gg v_*(t)$, while within the boundary
layer, $v\ll v_*(t)$, the velocity distribution continues to evolve.
The threshold velocity $v_*(t)$ is estimated by evaluating the first,
leading term in the right-hand side of Eq.~(\ref{asymp}): $t^{-1}P\sim
v_*^{\mu+1} P$, which implies $v_*\sim t^{-1/(\mu+1)}$.  The width of
the boundary layer shrinks with time but the boundary layer still exists
{\it ad infinitum}.

\begin{figure}
  \centerline{\epsfxsize=7.5cm \epsfbox{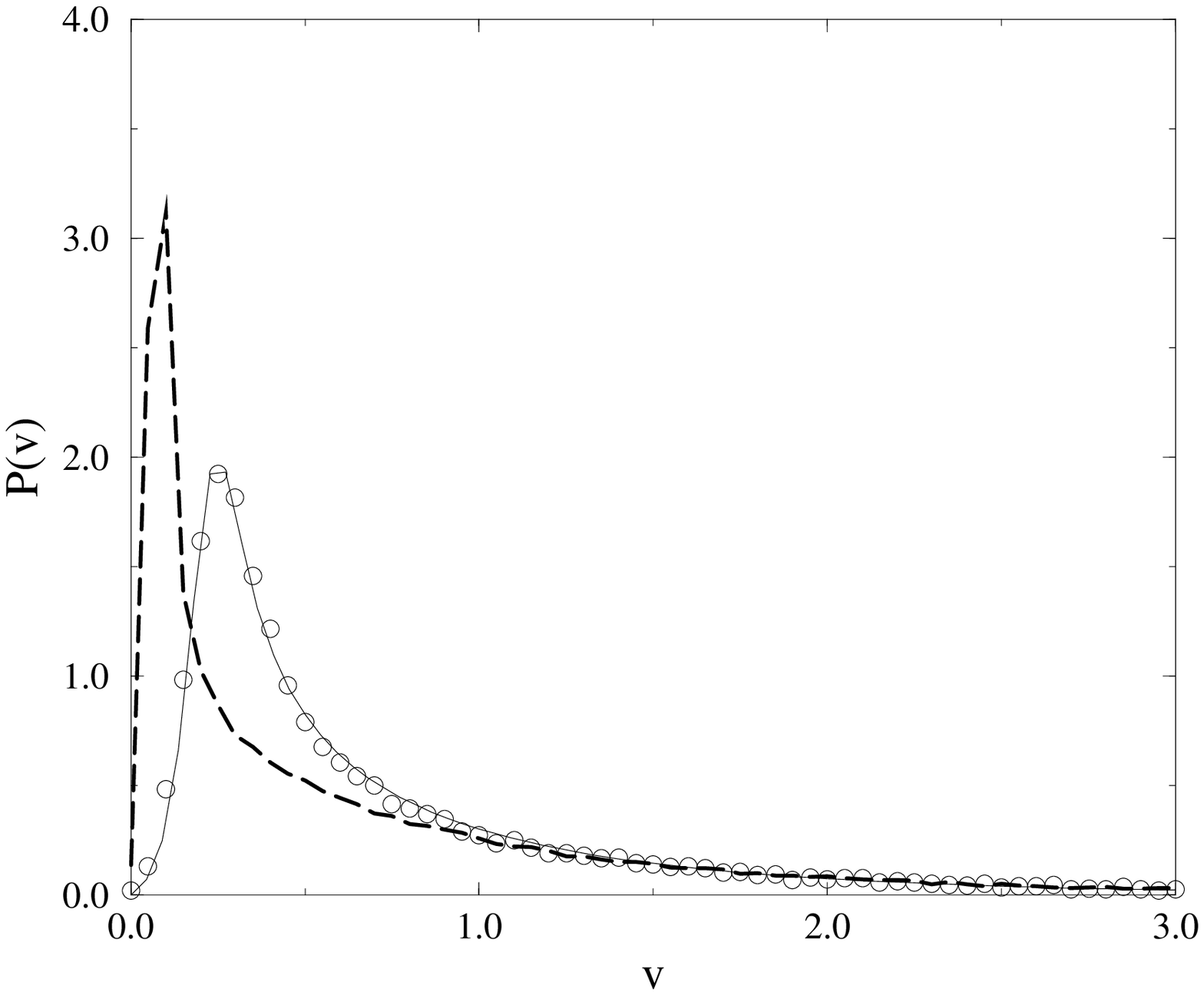}}
\noindent
{\small {\bf Fig.~1}.  Plot of $P(v,t)$ at 
  $t=200$ for $P_0(v)=v\,e^{-v}$: simulation result $(\circ)$, 
  numerical solution (---).  The dashed line shows simulation result for
  $P(v,t)$ at $t=16000$.}
\end{figure} 

Figs.~2--4 plot the average velocity vs. time for the intrinsic  
velocity distributions (\ref{p_0}) with $\mu=1, 0, -1/2$, respectively.
We find good agreement with the theoretical prediction of Eq.~(\ref{av})
when $\mu\geq 0$. On Fig.~5  the plot
of the local exponent $\alpha(t) \equiv d\ln[u(t)] / d \ln[t]$
vs. $t^{-1/2}$ is shown for $\mu=-1/2$. 
The results of extrapolation of $\alpha(t)$ to the 
$t\rightarrow \infty$ limit are not contradicting the
theoretical prediction, $\alpha=1/2$.

\begin{figure}
  \centerline{\epsfxsize=7.5cm \epsfbox{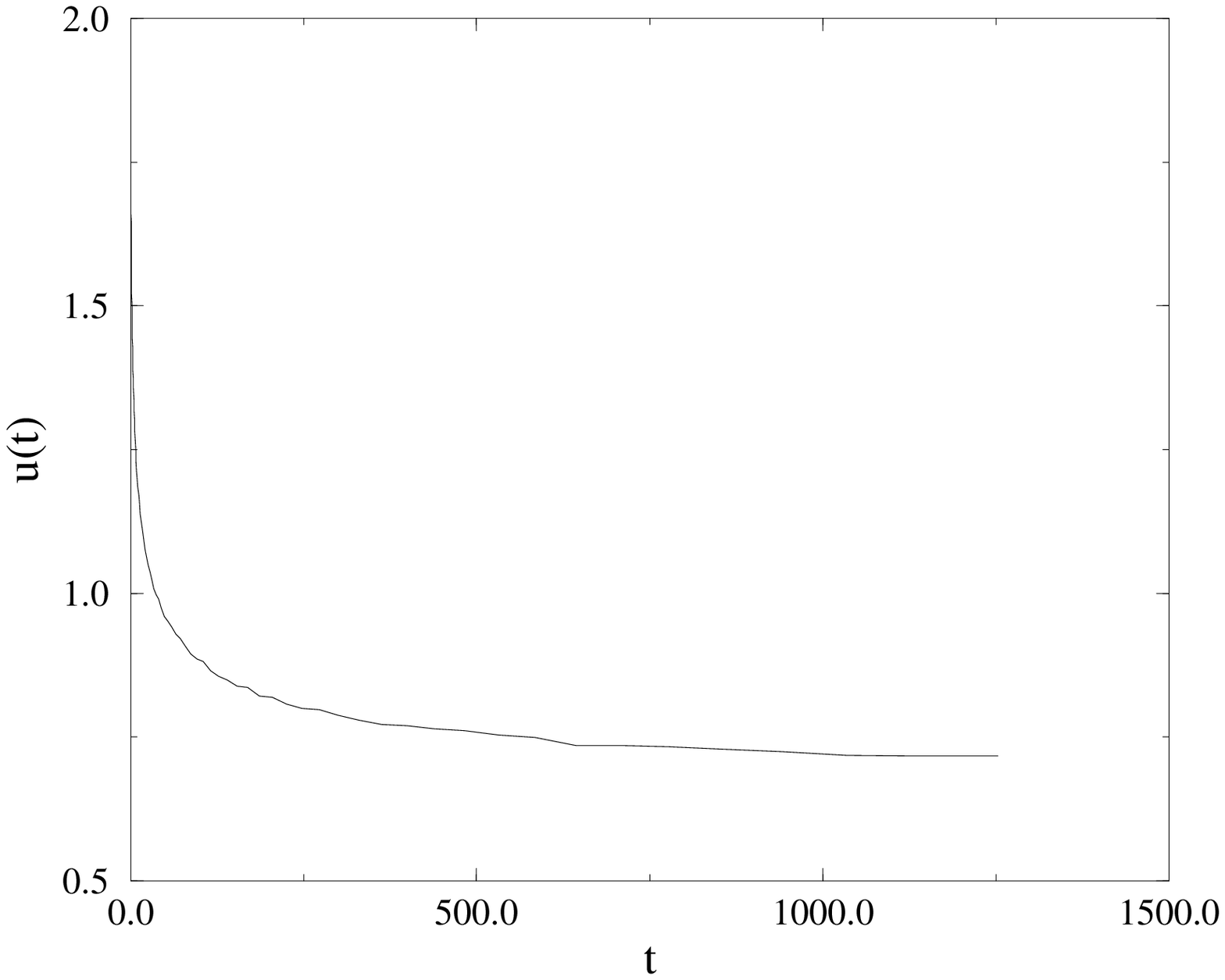}}
\noindent
{\small {\bf Fig.~2}.  Plot of $u(t)$ vs. time $t$ for
  $P_0(v)=v\,e^{-v}$.}
\end{figure} 
\begin{figure}
  \centerline{\epsfxsize=7.5cm \epsfbox{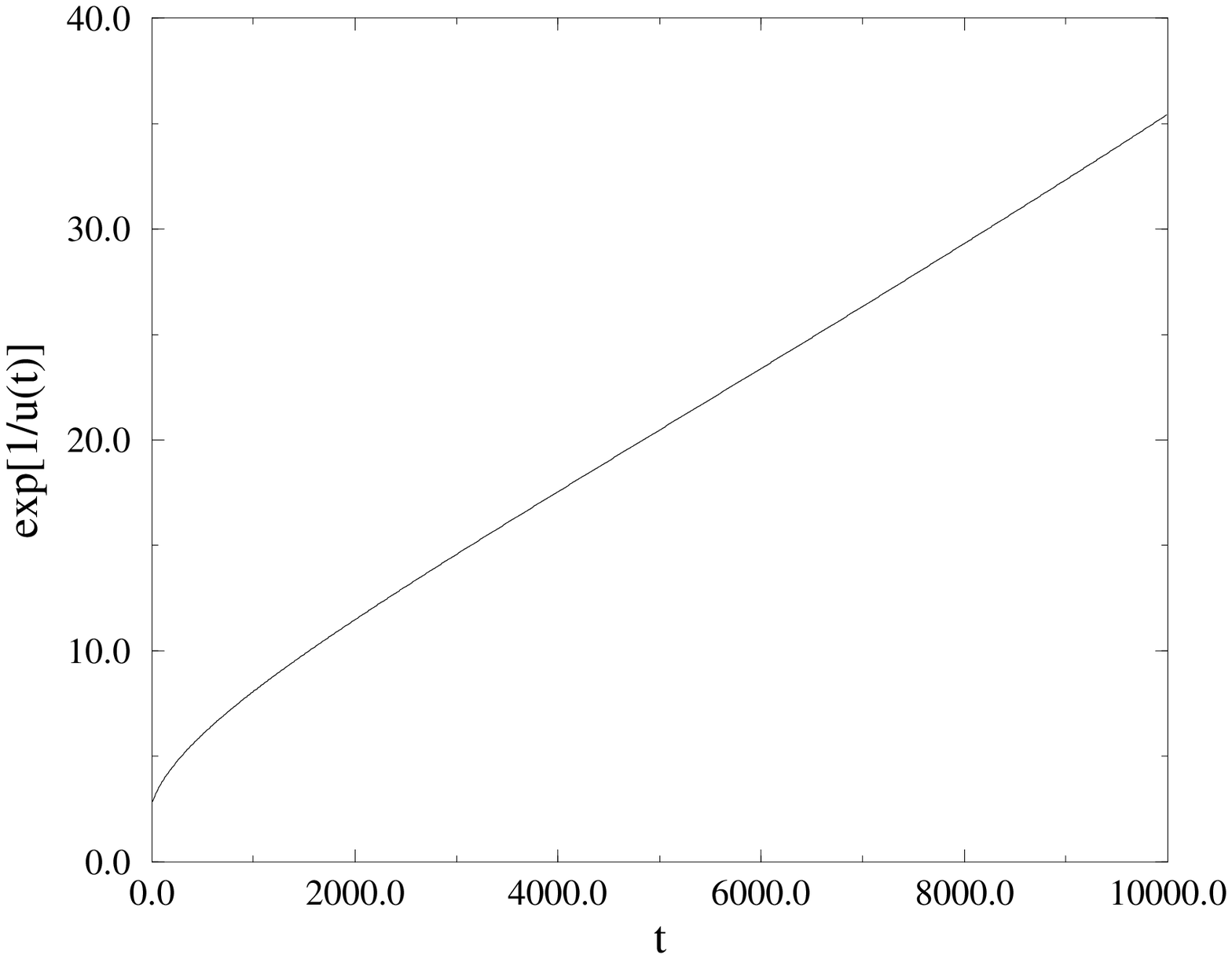}}
\noindent
{\small {\bf Fig.~3}.  Plot of $\exp[1/u(t)]$ vs. time $t$ for
  $P_0(v)=e^{-v}$.}
\end{figure} 
\begin{figure}
  \centerline{\epsfxsize=7.5cm \epsfbox{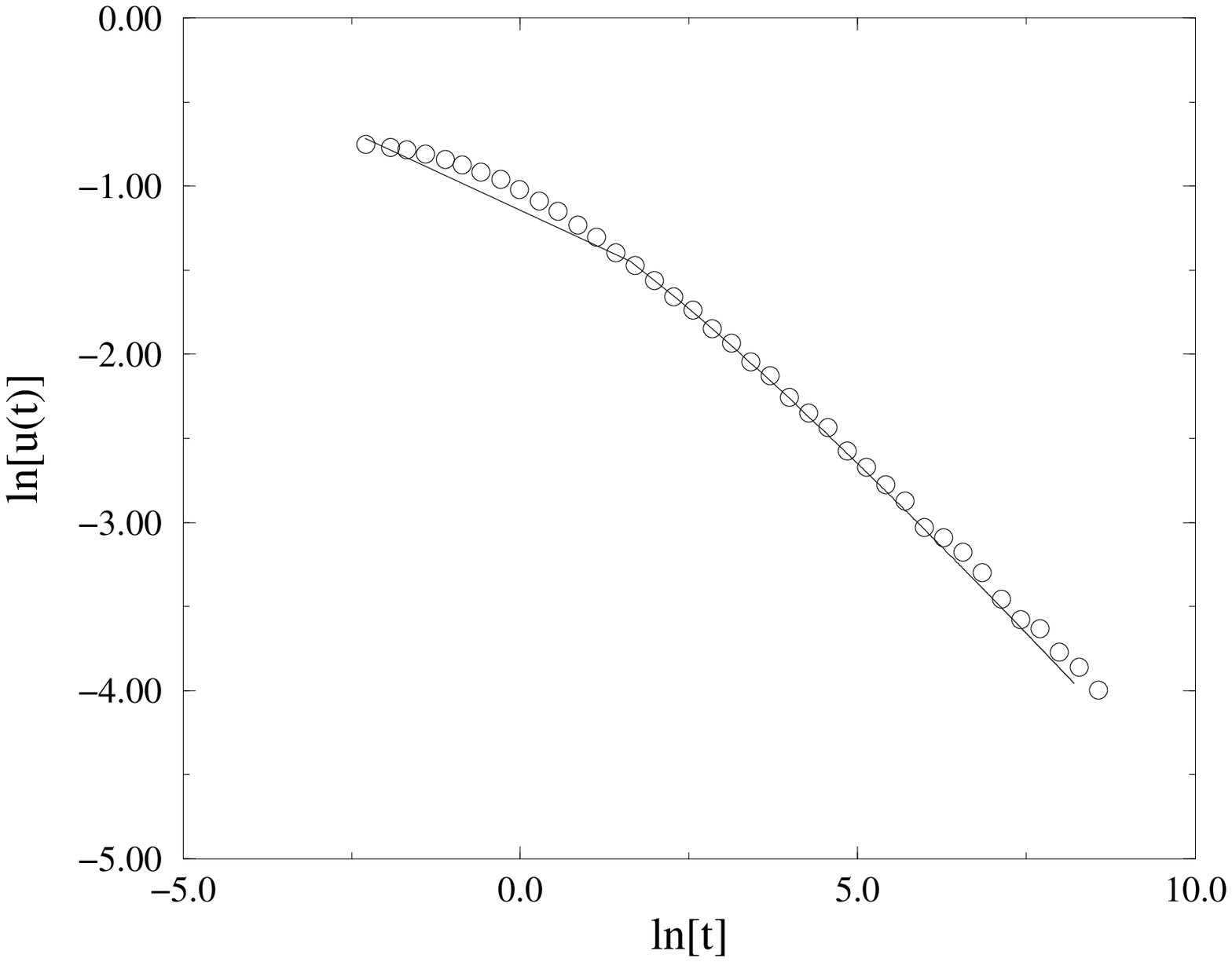}}
\noindent
{\small {\bf Fig.~4}.  Log-log plot of $u(t)$ vs. time
  $t$ for $P_0(v)=(\pi v)^{-1/2}e^{-v}$: molecular dynamics results
  $(\circ)$ and numerical solution (---)}.
\end{figure}
\begin{figure}
  \centerline{\epsfxsize=7.5cm \epsfbox{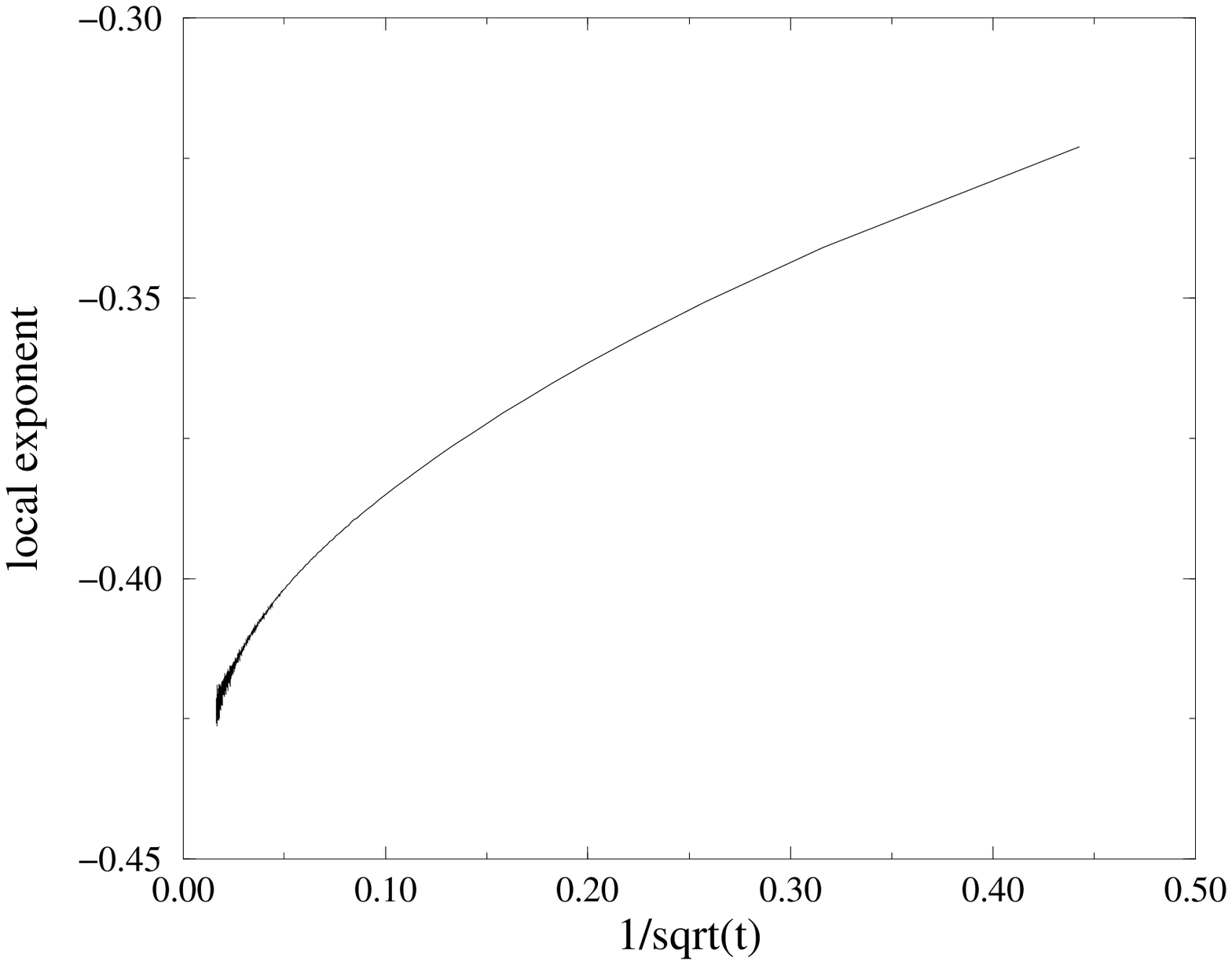}}
\noindent
{\small {\bf Fig.~5}.  Plot of the local exponent $\alpha(t)$ vs.
  $1/\sqrt{t}$ for $P_0(v)=(\pi v)^{-1/2}e^{-v}$.}
\end{figure} 

In summary, we have shown that the fate of the system of passing particles 
is determined by
the behavior of the intrinsic velocity distribution near its lower cutoff:
If $P_0(v)$ vanishes in this limit, the system reaches a
steady state; otherwise, the evolution continues forever.
Comparison between solutions of the mean-field Boltzmann equation and
results of molecular dynamics simulations suggests that the mean-field
theory description is exact.  It will be interesting to confirm this
result rigorously. 

\medskip\noindent
We gratefully acknowledge partial support from NSF and ARO.

\end{multicols}
\end{document}